\documentclass{article}
\usepackage{spconf,amsmath,graphicx}
\usepackage{tikz}
\usetikzlibrary{arrows,shapes,snakes,positioning,fit,backgrounds,mindmap,calc}
\usepackage[customcolors]{hf-tikz}

\tikzset{
    >=stealth',
    node/.style={
           rectangle,
           rounded corners,
           draw=black!80, very thick,
           text centered
    },
    env/.style={
           fill=blue!20,
    },
    tf/.style={
           fill=orange!60, 
    },
    wf/.style={
           fill=blue!40
    },
    pil/.style={
           ->,
           thick,
           shorten <=2pt,
           shorten >=2pt},
    inactive/.style={
          opacity=.3,
          dotted
    },
    inactive-pipeline/.style={
          opacity=.2
    }
}

\title{Estimating the ultrasound attenuation coefficient using convolutional neural networks -- a feasibility study}
\name{Piotr Jarosik$^{\dagger\ast}$  \qquad Michal Byra$^{\dagger}$  \qquad Marcin Lewandowski$^{\dagger}$  \qquad Ziemowit Klimonda$^{\dagger}$}
\address{
$^{\dagger}$Institute of Fundamental Technological Research, \\Polish Academy of Sciences, Warsaw, Poland\\
$^{\ast}$Corresponding author, e-mail: pjarosik@ippt.pan.pl}
\begin{document}
\nocite{matplotlib207hunter}

%
\maketitle
\begin{abstract}
Attenuation coefficient (AC) is a fundamental measure of tissue acoustical properties, which can be used in medical diagnostics. In this work, we investigate the feasibility of using convolutional neural networks (CNNs) to directly estimate AC from radio-frequency (RF) ultrasound signals. To develop the CNNs we used RF signals collected from tissue mimicking numerical phantoms for the AC values in a range from 0.1 to 1.5 dB/(MHz*cm). The models were trained based on 1-D patches of RF data. We obtained mean absolute AC estimation errors of 0.08, 0.12, 0.20, 0.25 for the patch lengths: 10 mm, 5 mm, 2 mm and 1 mm, respectively. We explain the performance of the model by visualizing the frequency content associated with convolutional filters. Our study presents that the AC can be calculated using deep learning, and the weights of the CNNs can have physical interpretation. 

\end{abstract}
\section{Introduction}
\label{sec:intro}

Ultrasound imaging is a popular medical imaging modality widely used in clinics. This modality is non-invasive, inexpensive and portable. However, in comparison to other modalities, such as computed tomography or magnetic resonance imaging, ultrasound images of human tissues are of relatively lower quality, making them more difficult to interpret by the radiologists. Various computer-aided diagnosis (CAD) systems have been developed to help the radiologists assess ultrasound images \cite{shiraishi2008computer, jalalian2013computer, flores2015improving}. Recently, we can observe an increasing interest in incorporating deep learning techniques into CAD systems \cite{yap2018automated,qi2019automated, jarosik2020breast}. Currently, the majority of the research is focused on developing solutions based on ultrasound B-mode images reconstructed using radio-frequency (RF) backscattered ultrasound signals.  Due to the reconstruction, however, information about RF signal's spectrum and phase is partially removed in order to make the output ultrasound image human-readable \cite{szabo2004diagnostic}. The lost frequency content may contain useful information about the examined structure.

Attenuation coefficient (AC) is one of the basic quantitative acoustic properties of human tissues. AC can be utilized for medical diagnosis, and has been used to differentiate liver \cite{lu1999ultrasound, kuc1980clinical} and breast \cite{d1986frequency} tissue pathologies. AC is commonly estimated based on RF signal's spectrum, by tracking the signals frequency content change with depth. For example, the mean frequency slope can be used to calculate the coefficient \cite{kuc1979estimating}. However the accuracy of the established methods may be disturbed by several factors, including the impact of transducer's characteristics and electrical noise.

In this work, we experimentally investigate a deep learning based approach to the AC estimation. We use RF signals to train convolutional neural networks (CNNs) for the direct AC calculations.  In our feasibility study, we verify model's performance depending on the amount of RF data provided to the CNN. We also visualize its internal representations to verify if any information related to the expected change in signal frequency content can be discovered.  

Deep convolutional networks have been already successfully used for the processing of raw acoustic signals -- e.g. in automatic speech recognition \cite{sainath2015learning, sainath2015convolutional, golik2015convolutional, hoshen2015speech}. In particular, Sainath et al. presented that convolutional layers, when properly trained, can learn to do a spectral filtering of the input signal. Our paper extends prior work by verifying (1) if the signal frequency content loss, specific for higher ultrasound attenuation, is truly represented in the neural network's weights and (2) what is the CNN's performance depending on the amount of the input data. The purpose of the first point is to increase CNN's interpretability, what is of great importance in medical applications. The answer to the second point will show what is the possible output resolution of the proposed method.

\section{Method}

\begin{figure}[t!]

\begin{minipage}[b]{1\linewidth}

\begin{minipage}[b]{.48\linewidth}
  \centering
  \centerline{\includegraphics[width=1\linewidth]{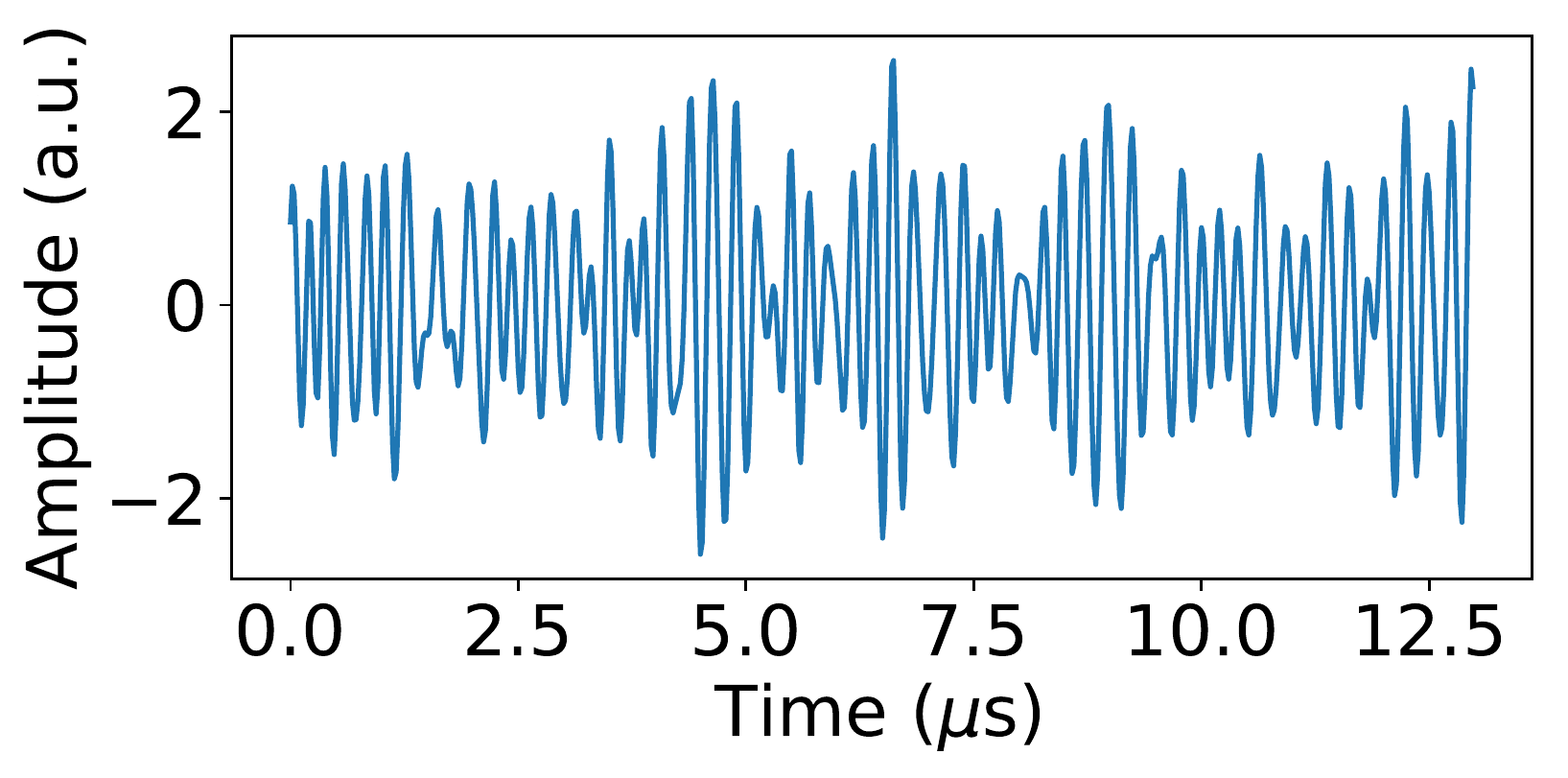}}
\end{minipage}
\hfill
\begin{minipage}[b]{.48\linewidth}
  \centering
  \centerline{\includegraphics[width=1\linewidth]{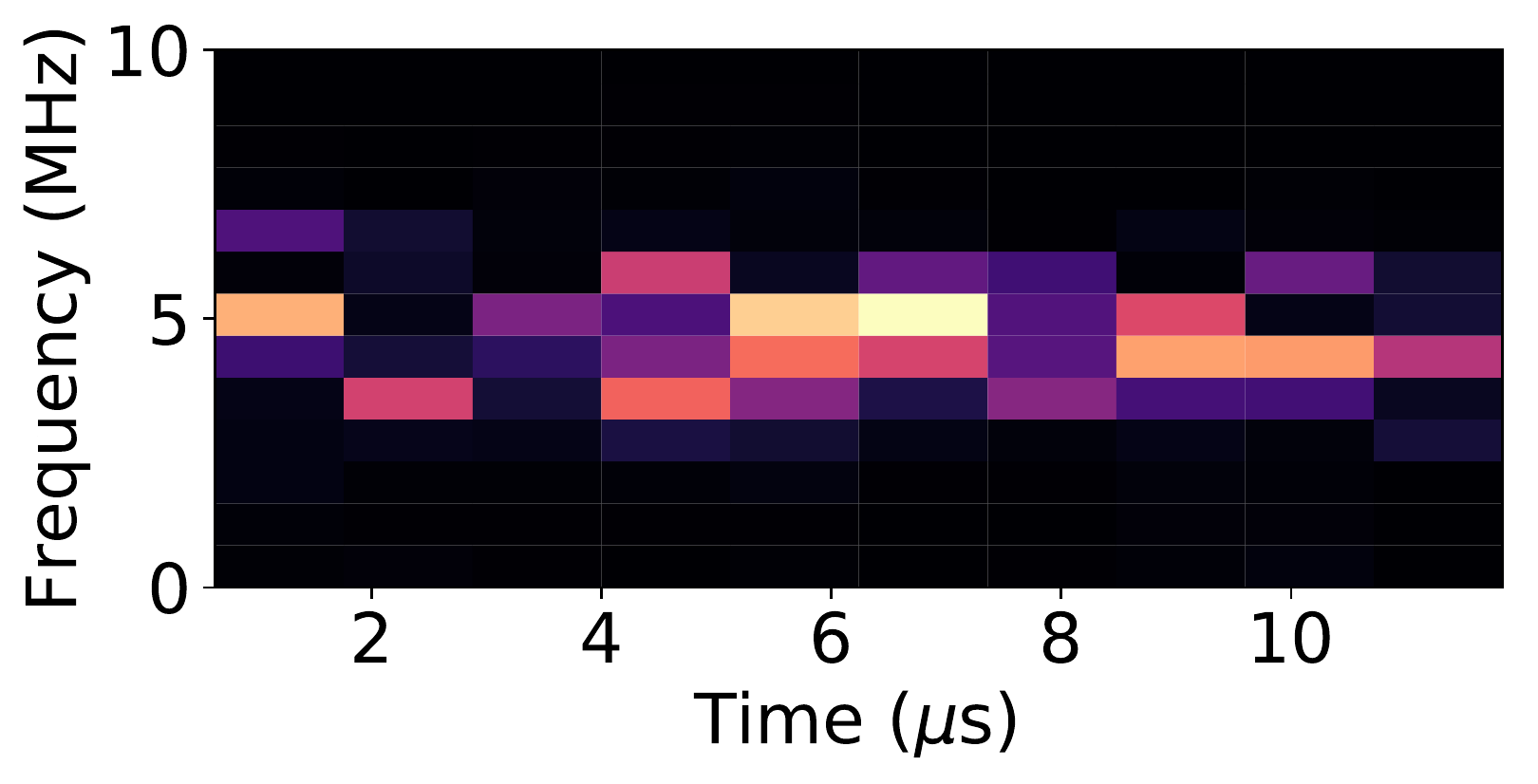}}
\end{minipage}
  \centerline{(a)}\medskip
\end{minipage}
\begin{minipage}[b]{1\linewidth}
\begin{minipage}[b]{.48\linewidth}
  \centering
  \centerline{\includegraphics[width=1\linewidth]{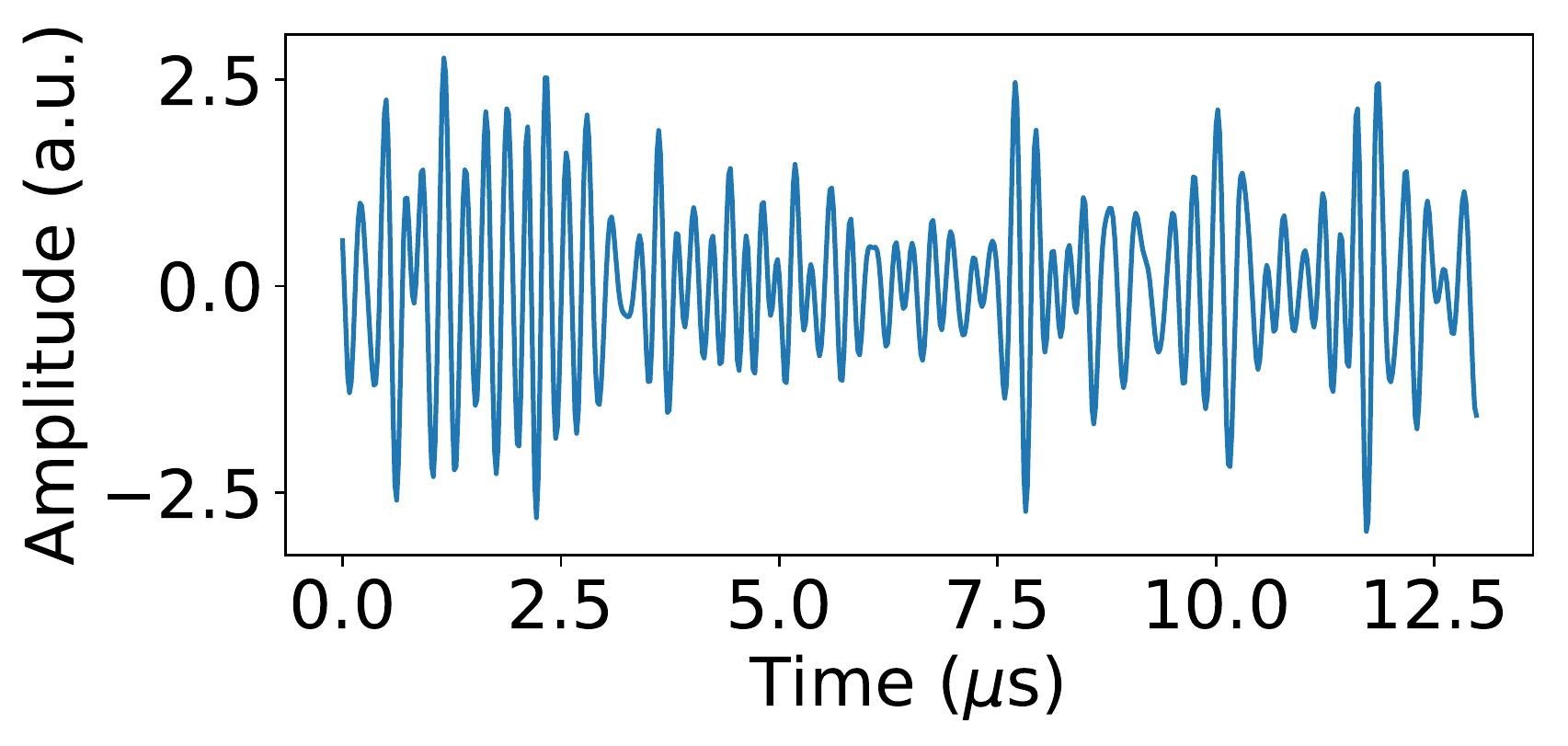}}
\end{minipage}
\hfill
\begin{minipage}[b]{.48\linewidth}
  \centering
  \centerline{\includegraphics[width=1\linewidth]{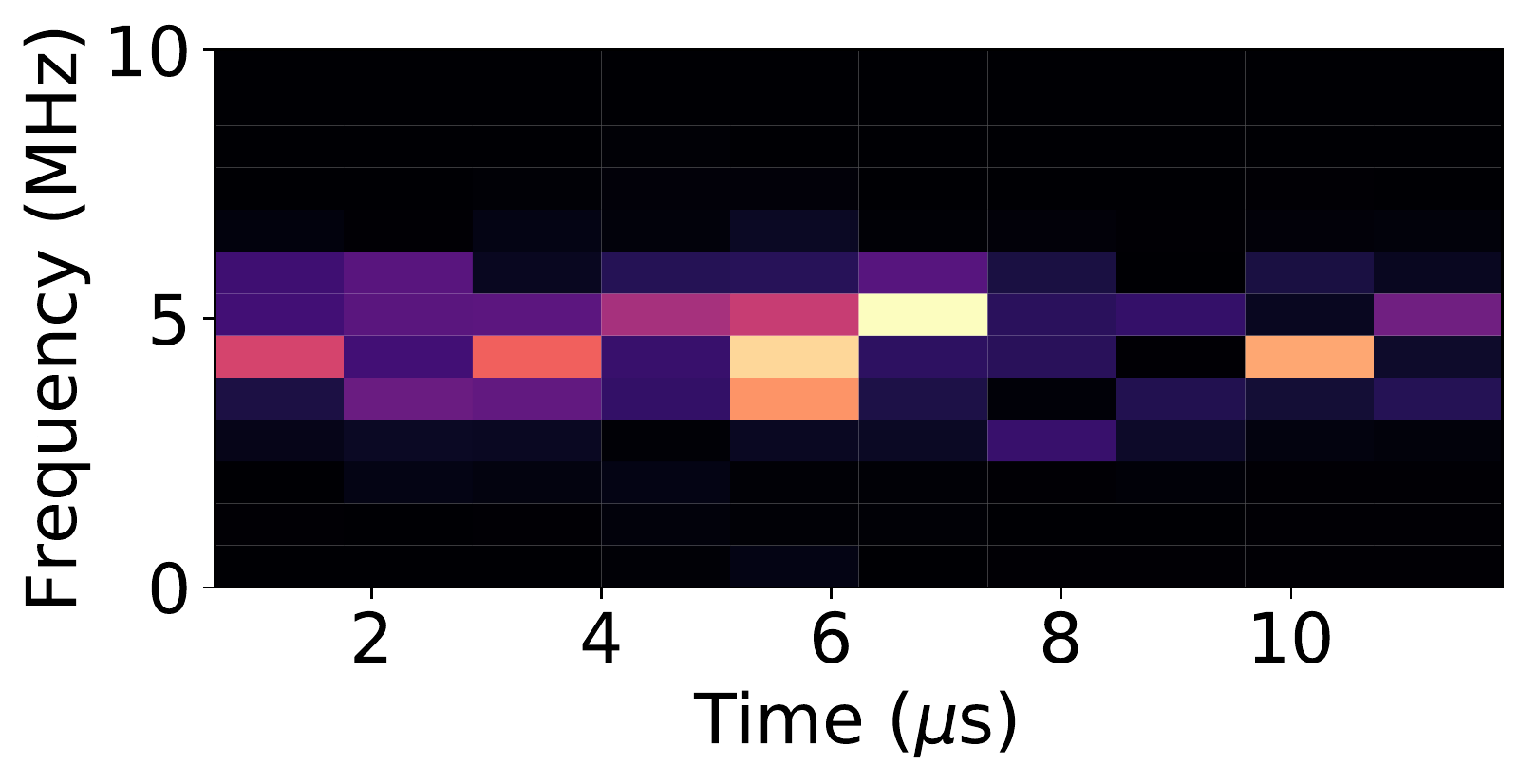}}
\end{minipage}
\centerline{(b)}\medskip
\end{minipage}
\begin{minipage}[b]{1\linewidth}
\begin{minipage}[b]{.48\linewidth}
  \centering
  \centerline{\includegraphics[width=1\linewidth]{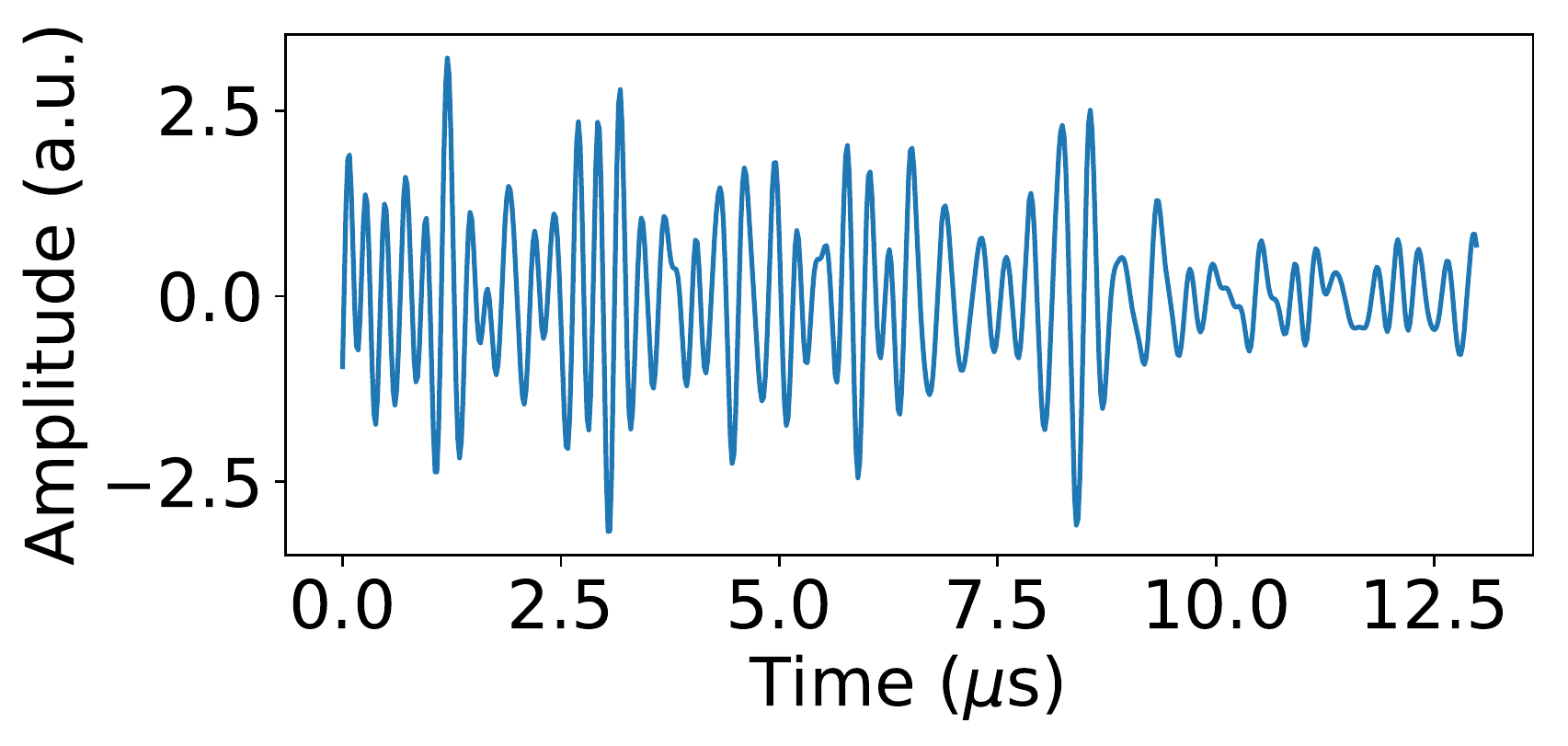}}
\end{minipage}
\hfill
\begin{minipage}[b]{.48\linewidth}
  \centering
  \centerline{\includegraphics[width=1\linewidth]{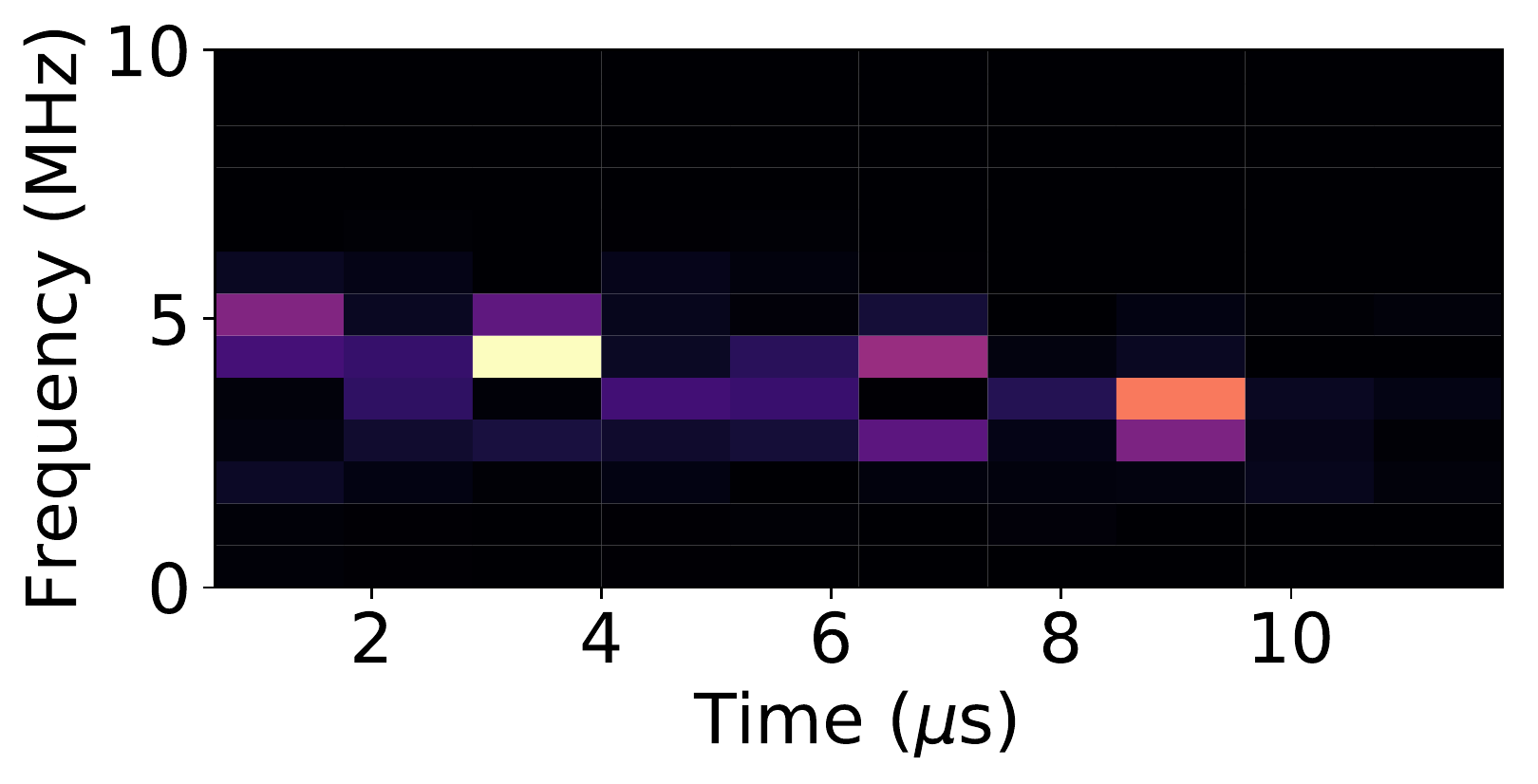}}
\end{minipage}
\centerline{(c)}\medskip
\end{minipage}
\caption{Examples of 10 mm RF data chunks (waveform, spectrogram), which were used in our experiments. The data were acquired from simulation of ultrasound scattering in a soft tissue with AC equal to (a)~0.1 (b)~0.7 (c)~1.5. High-frequency components are more strongly suppressed by the structures with higher AC value.}
\label{fig:dataset}
\end{figure}

\subsection{Dataset}
 We used Field-II software (created by Jensen et al. \cite{jensen1992calculation, jensen1996field}) to simulate an ultrasound wave propagation and to generate 1024 RF lines for each attenuation level $\{0.1, 0.2, ..., 1.5\}$ dB/(MHz*cm), 15360 lines in total. A piston transducer with a diameter of $d_t = 12$ (mm) and center frequency $f_0 = 5$ (MHz) was used. Backscattered echo signal was digitized with $f_s = 50$ (MHz) sampling rate. An  ultrasound impulse was propagating through tissue mimicking medium with a given attenuation level. The maximum depth of signal acquisition was equal to 50 (mm). A speed of sound with value $c = 1540$ (m/s) was set. A digitized RF scanline consisted of approximately 3400 samples.

We applied the sliding window technique to split RF lines into multiple smaller fragments (1-D patches). A rectangular window of length $k \in \{1, 2, 5, 10\}$ (mm) was used. Each patch was normalized by subtracting its mean and dividing by its standard deviation. Next, processed data were used to train deep learning models.

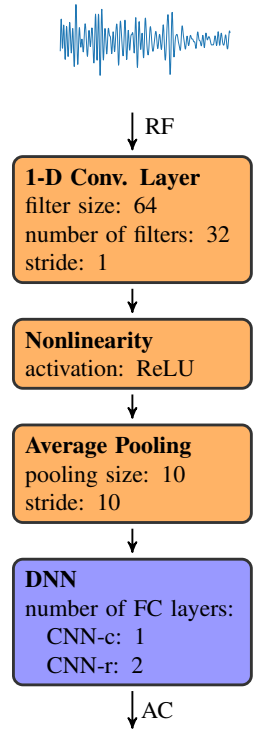
\begin{figure}
\centering
\scalebox{.9}{
\begin{tikzpicture}[->,node distance=1cm, auto,]
  \node[](env){};
  \node[anchor=south east,inner sep=0pt] at ($(env.south east)-(-.5cm,.4cm)$) {RF};
  \node[anchor=south east,inner sep=0pt] at ($(env.south east)-(-1.5cm,0cm)$) {
  \includegraphics[width=.4\linewidth]{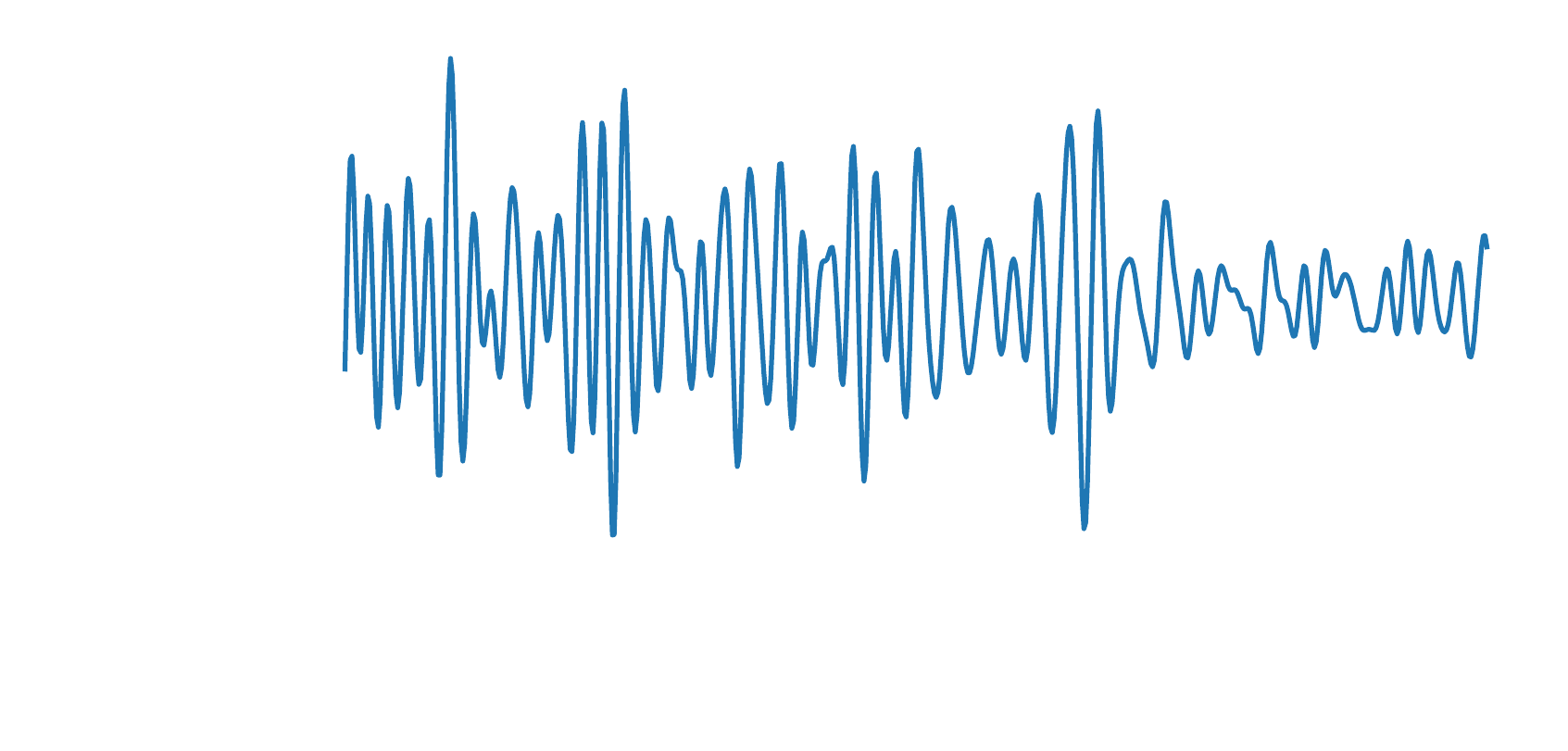}
  };
  \node[node, tf, inner sep=5pt,below=.7cm of env, align=left, text width=9em](bf){
  \textbf{1-D Conv. Layer}\\
  filter size: 64\\
  number of filters: 32\\
  stride: 1
  };
  \node[node, tf, inner sep=5pt,below=.5cm of bf, align=left, text width=9em](as){
  \textbf{Nonlinearity}\\
  activation: ReLU
  };
  \node[node, tf, inner sep=5pt,below=.5cm of as, align=left, text width=9em](abs){\textbf{Average Pooling}\\
  pooling size: 10\\
  stride: 10
  };
  \node[node, wf, inner sep=5pt,below=.5cm of abs, text width=5em, align=left, text width=9em](drc){
  \textbf{DNN}\\
  number of FC layers: \\
  \hspace{0.2cm} CNN-c: 1\\
  \hspace{0.2cm} CNN-r: 2
  };
  \node[draw=none] (end) [below=.7cm of drc] {};
  \path[every node/.style={transform shape, text centered}]
   (env) edge[pil] node [right, pos=0.1] {} (bf)
   (bf) edge[pil] node [above] {} (as)
   (as) edge[pil] node [above] {} (abs)
   (abs) edge[pil] node [above] {} (drc)
   (drc) edge[pil] node [right] {AC} (end);
\end{tikzpicture}
}
\caption{Neural network architecture evaluated in this work. DNN block consists of multiple fully connected (FC) layers.}
\label{fig:nn}
\end{figure}

\subsection{Models and evaluation procedure}
\begin{figure*}[t!]
\begin{minipage}[b]{1\linewidth}
\begin{minipage}[b]{.24\linewidth}
  \centering
  \centerline{\includegraphics[width=1\linewidth]{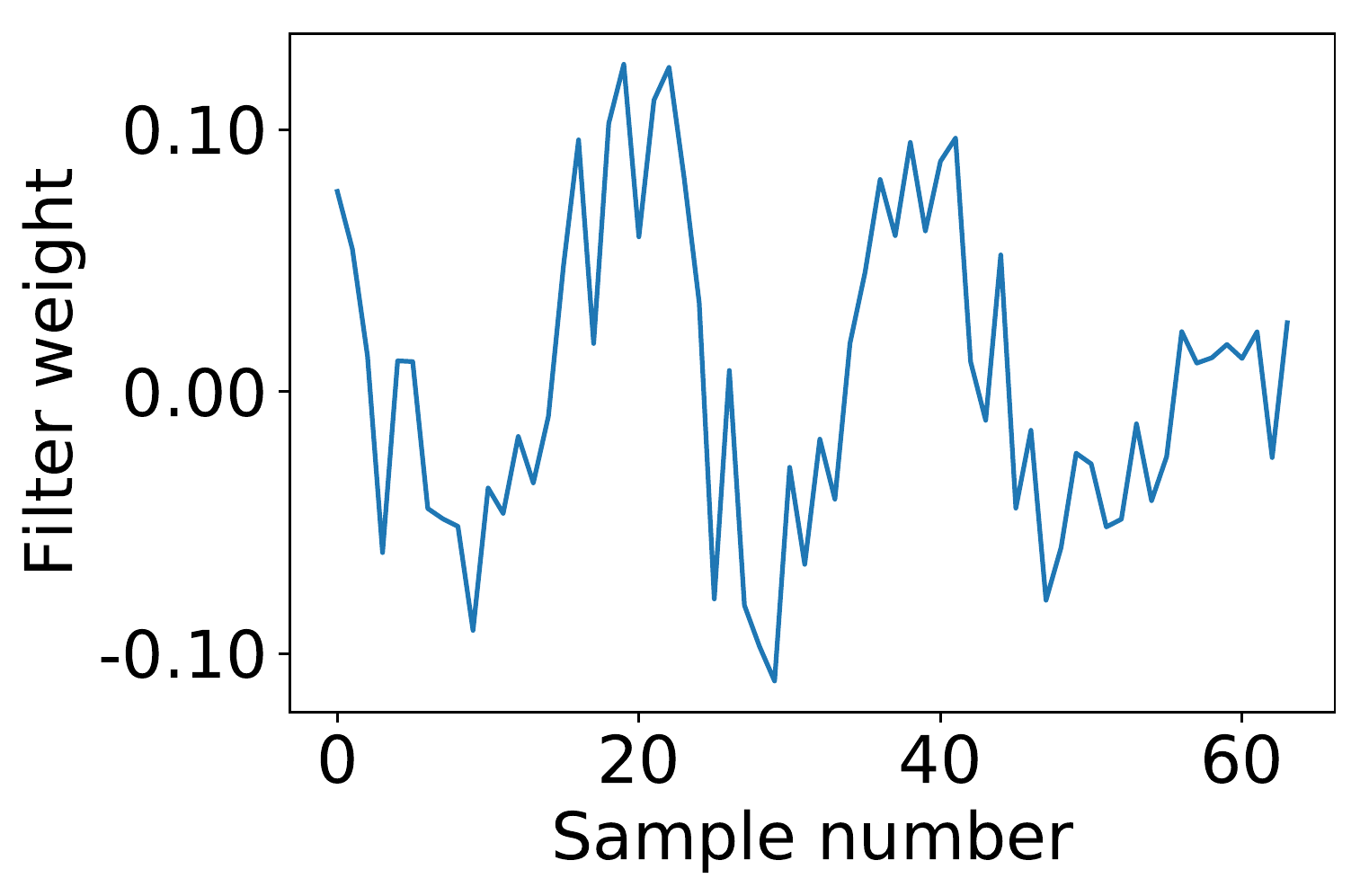}}
\end{minipage}
\hfill
\begin{minipage}[b]{.24\linewidth}
  \centering
  \centerline{\includegraphics[width=1\linewidth]{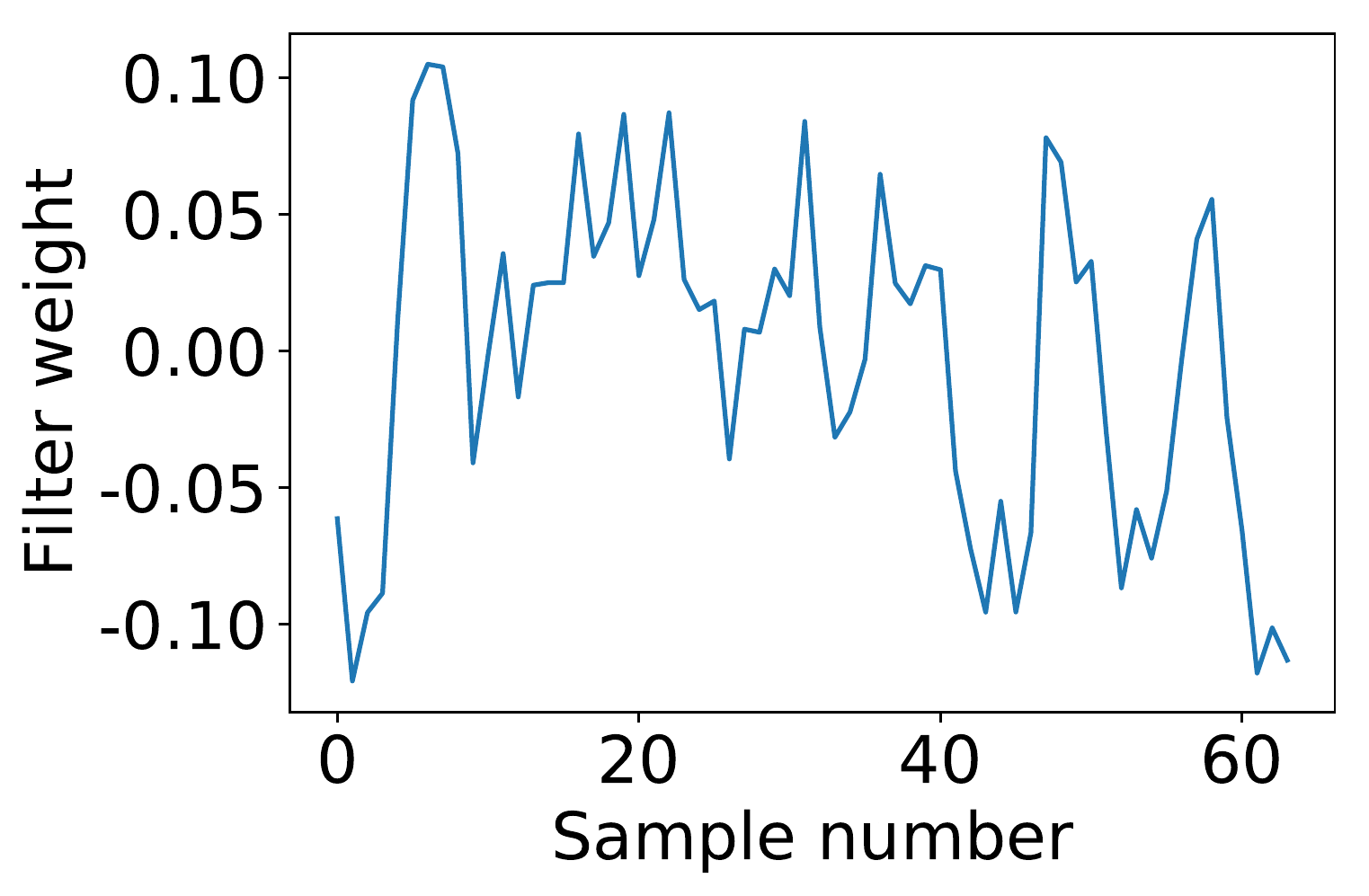}}
\end{minipage}
\begin{minipage}[b]{.24\linewidth}
  \centering
  \centerline{\includegraphics[width=1\linewidth]{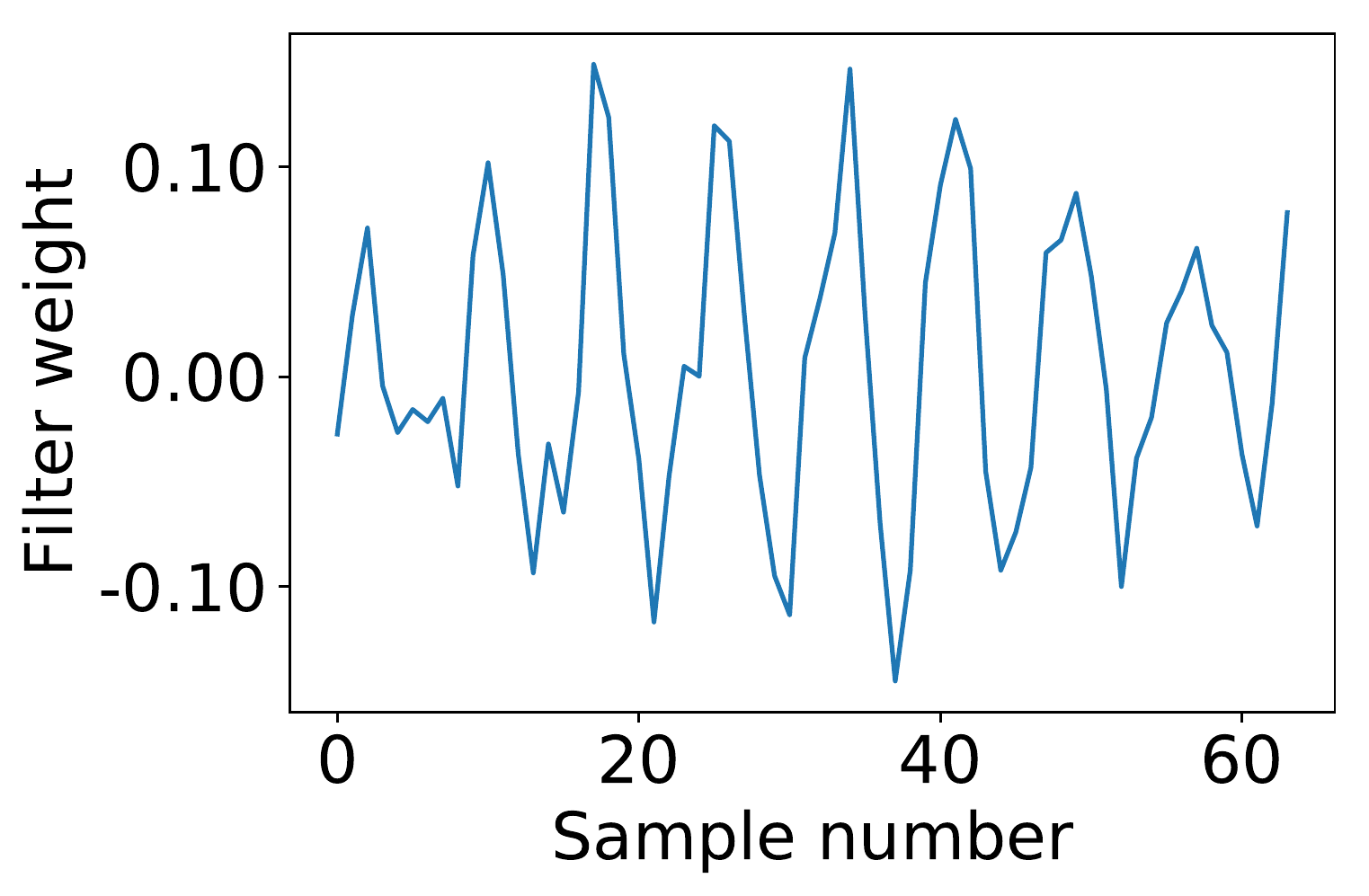}}
\end{minipage}
\hfill
\begin{minipage}[b]{.24\linewidth}
  \centering
  \centerline{\includegraphics[width=1\linewidth]{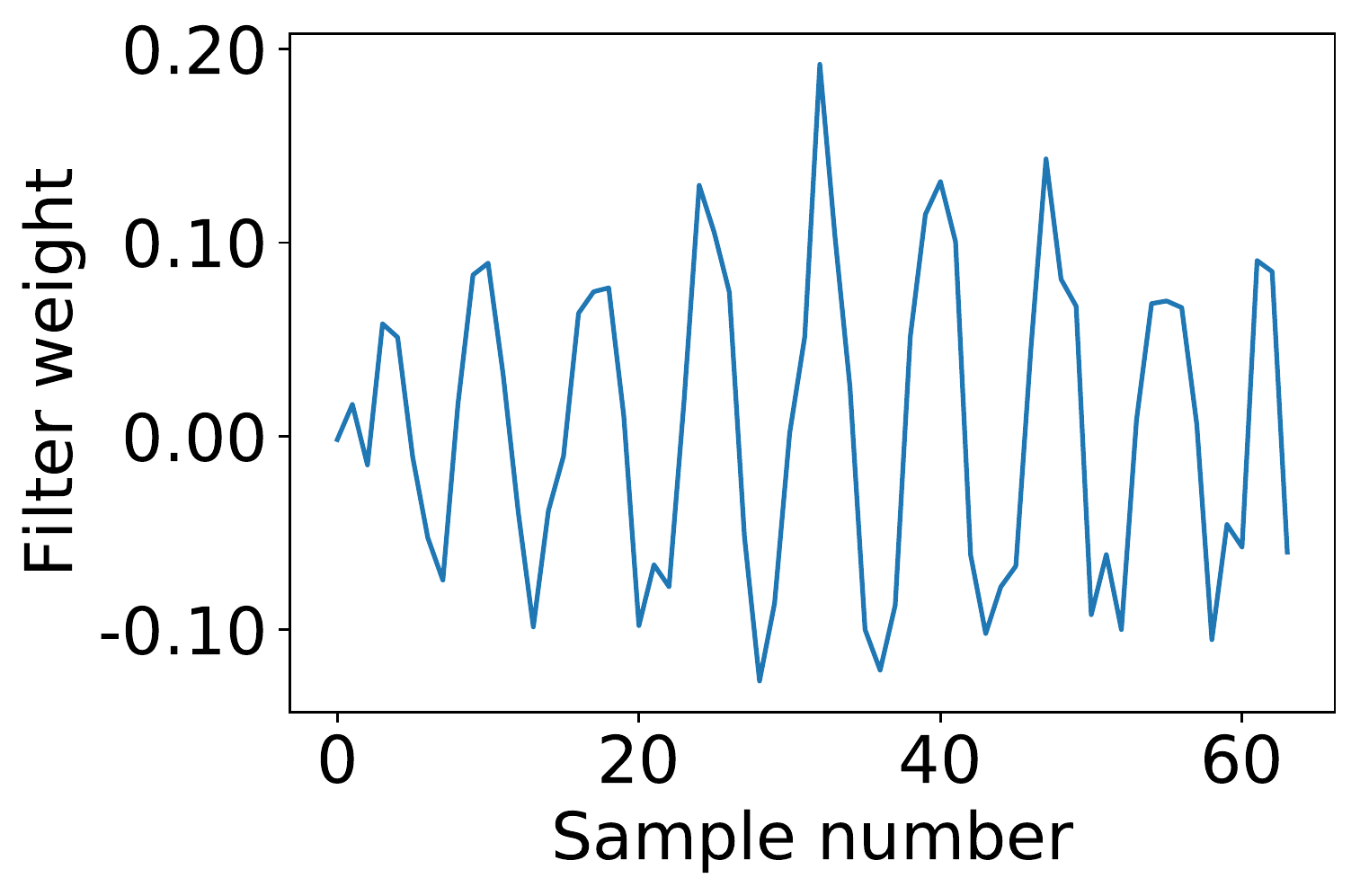}}
\end{minipage}
\centerline{(a)}\medskip
\end{minipage}
\begin{minipage}[b]{.48\linewidth}
  \centering
  \centerline{\includegraphics[width=.75\linewidth]{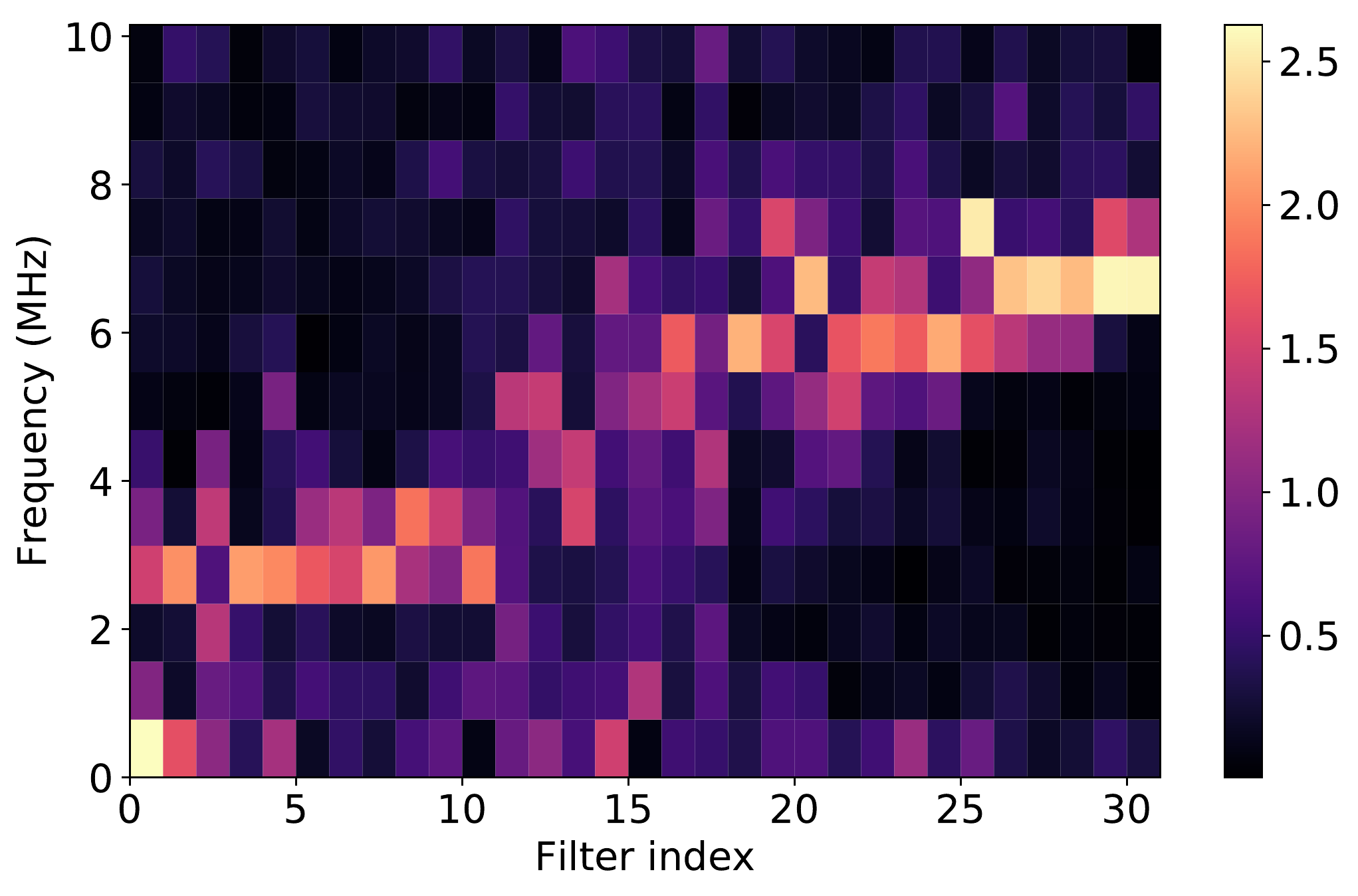}}
  \hspace*{0.1cm}
  \centerline{(b)}\medskip
\end{minipage}
\begin{minipage}[b]{.48\linewidth}
  \centering
  \centerline{\includegraphics[width=.75\linewidth]{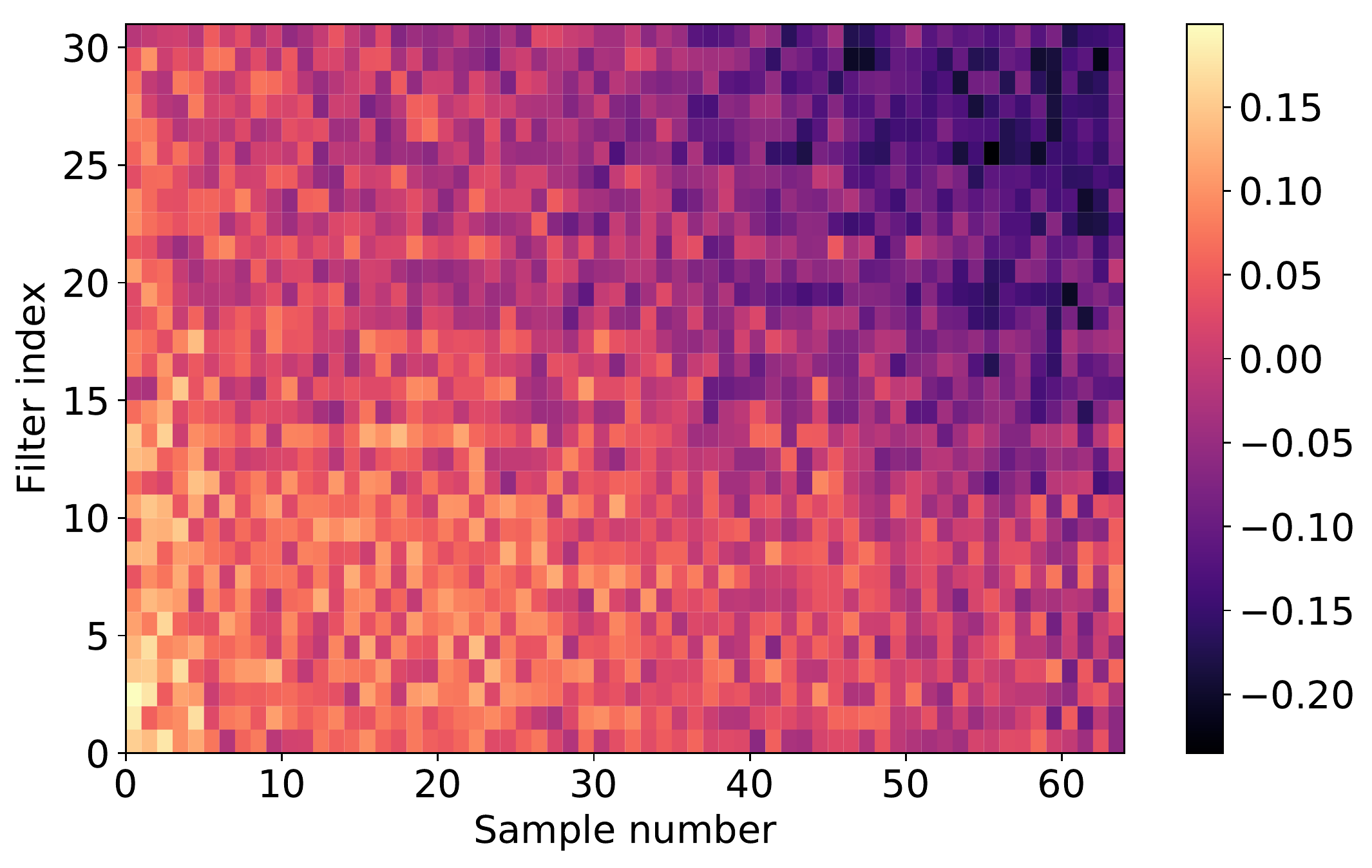}}
  \hspace*{0.1cm}
  \centerline{(c)}\medskip
\end{minipage}
\caption{Visualization of the CNN-c model developed to distinguish between AC values of 0.1 and 1.5. (a) Four sample filters of the first 1-D convolutional layer (waveform). (b). Magnitude spectrum of the first 1-D convolutional layer filters. Each column presents frequency content of one kernel. Filters were sorted according to the mean frequency. (c) Weights of the second, fully-connected layer, which was detecting AC = 1.5. The model conducted the classification decision based on the frequency content of input signals.}
\label{fig:cnn-r}
\end{figure*}

\label{sec:models}

We developed multi-layered artificial neural networks for the purposes of our experiments. Each neural model had a similar structure: input signal was processed by a 1-D convolutional layer followed by a 1-D average pooling layer, (see Fig. \ref{fig:nn}). The output was flattened, then processed by several fully-connected layers. Similar approaches have been applied for acoustic signal processing, for instance by Golik et al. \cite{golik2015convolutional} and Sainath et al. \cite{sainath2015convolutional}. We used ReLU activation function \cite{nair2010rectified} and applied batch normalization \cite{ioffe2015batch}. Neural network weights were initialized using Glorot technique \cite{glorot2011deep}.

We experimented with two neural network models to verify and interpret CNN's ability to distinguish attenuation levels and to estimate the AC value. The first architecture (named CNN-c) contained 1 fully connected layer, had a sigmoid output activation and was trained to discriminate attenuation levels 0.1 and 1.5 (a classification task) by minimizing binary cross-entropy. We attempted to interpret a knowledge discovered by a learning algorithm and hidden in the CNN-c parameters (see Fig. \ref{fig:cnn-r}). The second architecture (named CNN-r) contained 2 fully connected layers, had a ReLU output activation and was trained to estimate AC value (a regression task) by minimizing mean absolute error:
\begin{equation}
    \label{eq:mae}
    L = \frac{1}{N}\sum_{i=1}^{N}|y_i-\hat{y}_i| 
\end{equation}
where $\hat{y}_i$ is an estimated AC value and $y_i$ is a true coefficient value.
We used CNN-r to evaluate the method's performance depending on the size of the RF input data.

We used cross validation procedure to assess method's performance. We randomly split the dataset into train, validation and test subsets. We used 50\% of all RF lines for training, 20\% for validation and hyper-parameter tuning and 30\% for the final testing. We used Adam optimization algorithm with learning rate equal $10^{-3}$ to minimize the loss functions. We assessed model's performance using mean absolute error (Eq. \ref{eq:mae}) and standard deviation of the absolute errors $e = |y_i-\hat{y}_i|$.

\section{Results and Discussion}

\subsection{Interpretability}

Visualization of the CNN-c parameters is presented in Fig. \ref{fig:cnn-r}. First, several kernels of the 1-D convolutional layer resembled bandpass filters (both in terms of waveform and the spectrum). Similar after-training observations were reported for auditory-like filters in related publications \cite{sainath2015convolutional, golik2015convolutional}. Each kernel had a mean frequency located in the range [0, 10] (MHz), most were close to 5 (MHz). That conforms with dataset generation parameters: the center frequency of an ultrasound impulse was equal $f_0$ = 5 (MHz). 

Moreover, filters with the highest mean frequency (indices 22-31, $f_m \approx$ 7 (MHz)) have relatively narrow band. It is important to note here, that nonexistence of higher-frequency in the acquired ultrasound signal may be a good indicator, that the ultrasound signal comes from an area with sufficiently high attenuation. According to Kuc et al. \cite{kuc1980clinical}, some of the human tissues (like the liver), can "\emph{behave like a distributed acoustic low-pass filter}". The natural way is thus to expect, that this kind of information will be used by the appropriately trained model.

Our analysis conforms with the next observation: weights of the output fully-connected layer (which detects AC = 1.5) were negative for the output from kernels with $f_m \approx 7$ (MHz), approximately at the end of the processed 1-D patch. This means that the model performed AC detection based on the temporal changes of the spectrum. The greater the loss of high-frequency components, the greater probability that the attenuation was high. Thus the underlying CNN's operations behaved similarly to other state-of-the-art AC estimation methods. 

\begin{figure*}[t!]

\begin{minipage}[b]{.24\linewidth}
  \centering
  \centerline{\includegraphics[width=1\linewidth]{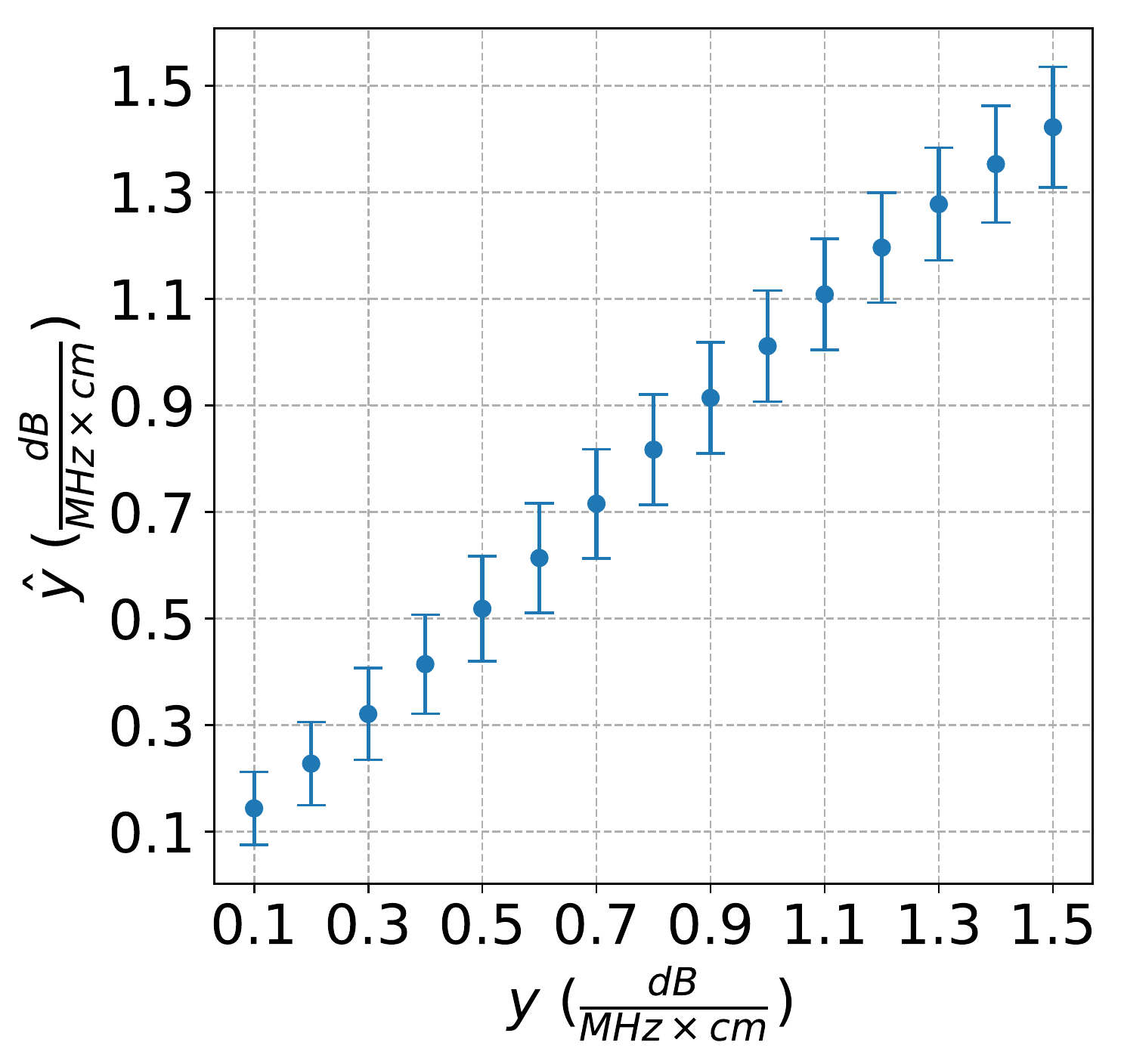}}
\hspace*{0.25cm}
  \centerline{k = 10 (mm)}\medskip
\end{minipage}
\begin{minipage}[b]{.24\linewidth}
  \centering
  \centerline{\includegraphics[width=1\linewidth]{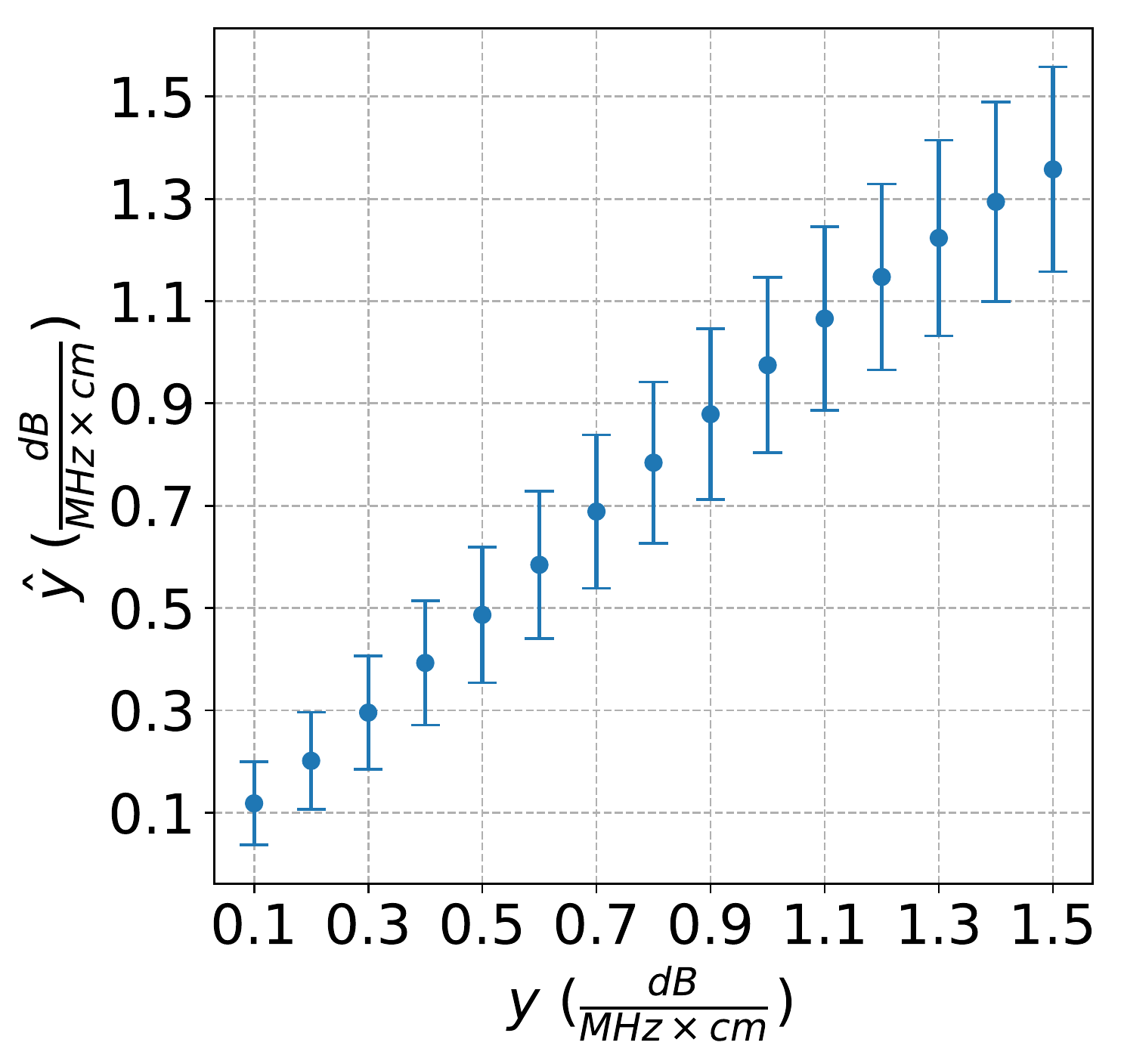}}
  \hspace*{0.25cm}
  \centerline{k = 5 (mm)}\medskip
\end{minipage}
\begin{minipage}[b]{.24\linewidth}
  \centering
  \centerline{\includegraphics[width=1\linewidth]{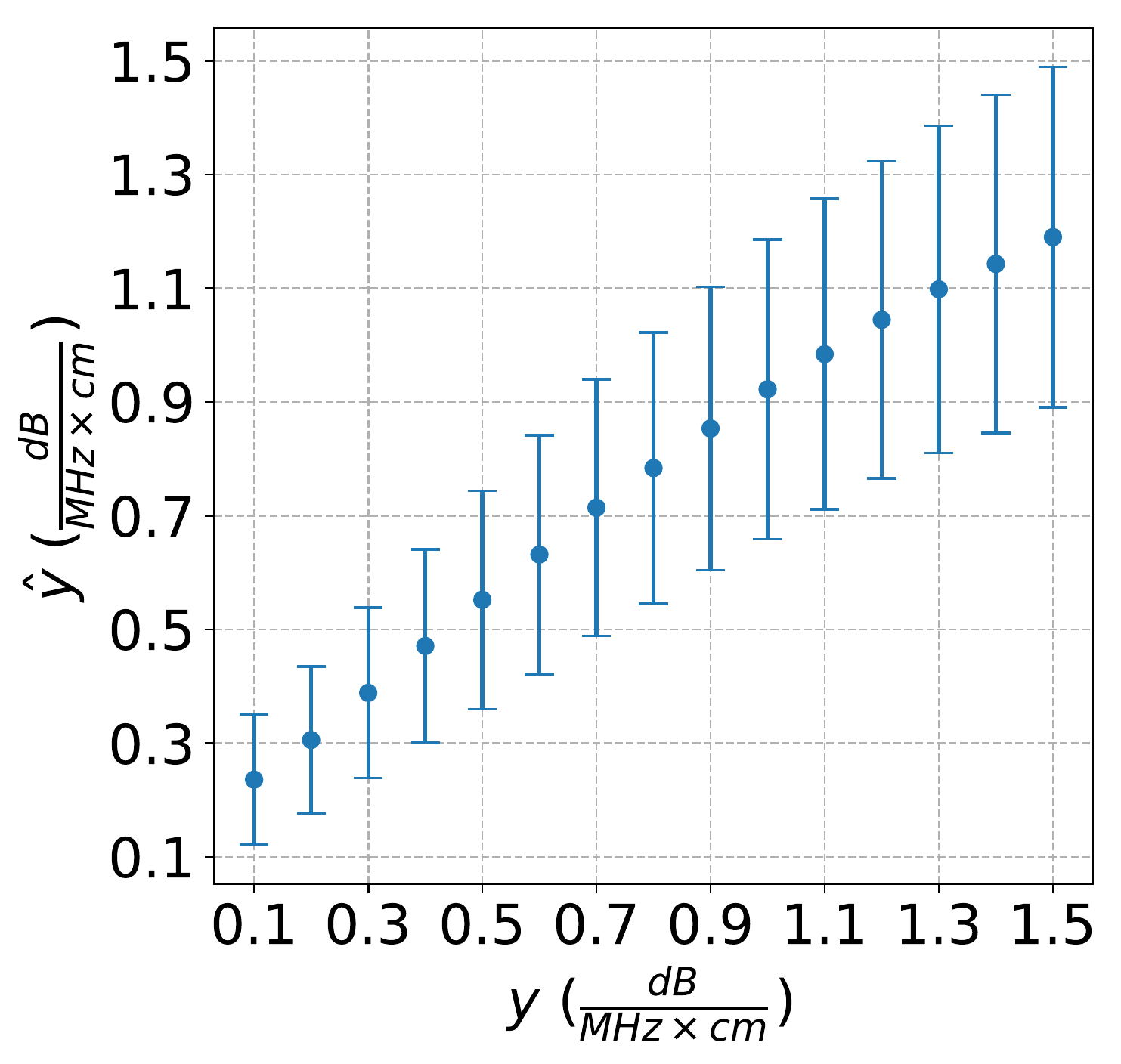}}
  \hspace*{0.25cm}
  \centerline{k = 2 (mm)}\medskip
\end{minipage}
\begin{minipage}[b]{.24\linewidth}
  \centering
  \centerline{\includegraphics[width=1\linewidth]{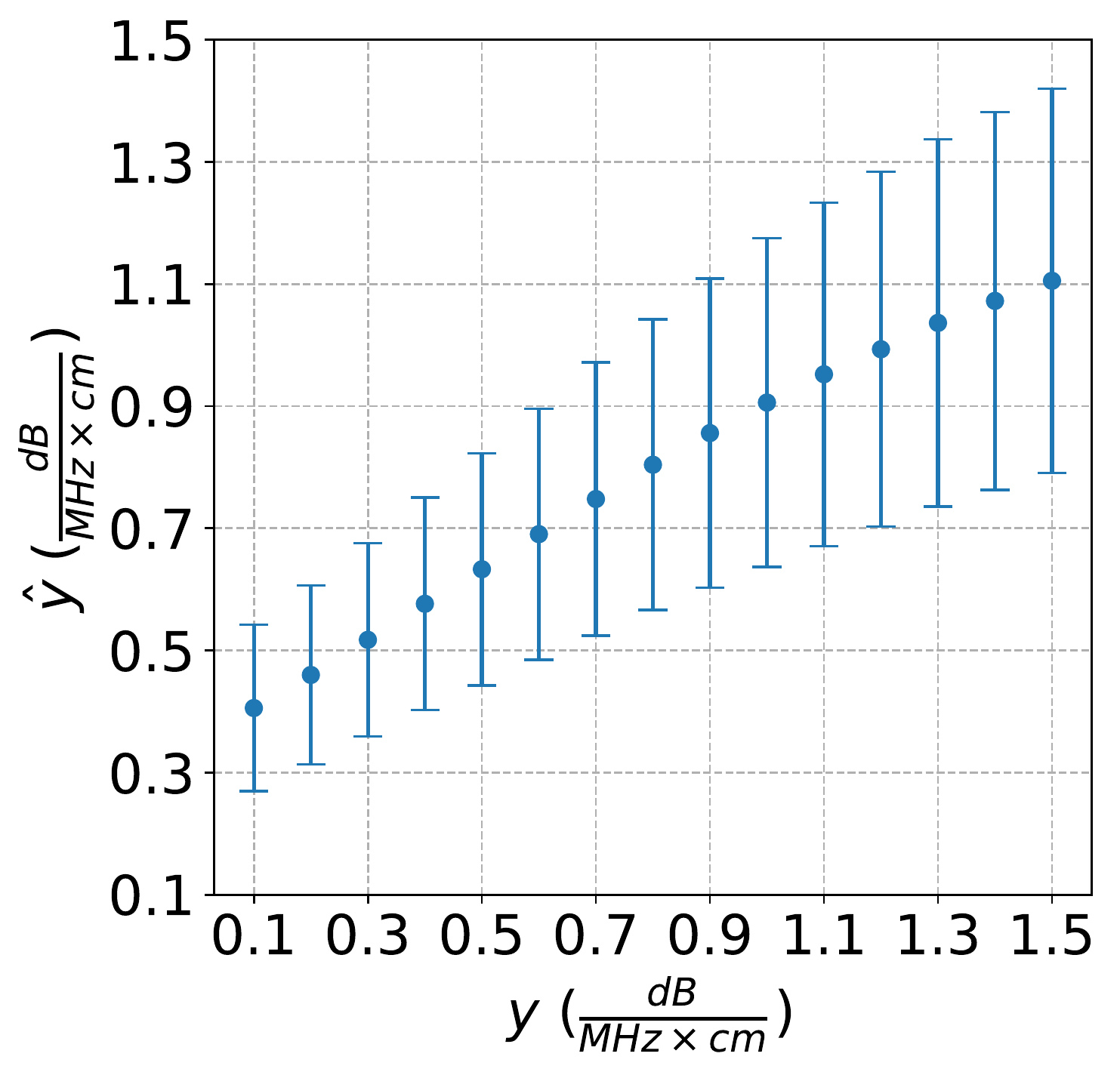}}
  \hspace*{0.25cm}
  \centerline{k = 1 (mm)}\medskip
\end{minipage}
\caption{Average AC estimate $\hat{y}$ computed by CNN based on RF echoes collected from tissue mimicking numerical phantoms with attenuation $y \in \{0.1, 0.2, ..., 1.5\}$. Each point represents  estimate's average $\hat{y}$; whiskers shows standard deviation range. The average AC estimate and standard deviation both increased with the sliding window size, equal to $k$.}
\label{fig:res}
\end{figure*}

\subsection{Performance}

Evaluation results are presented in Table \ref{tbl:results}. We obtained the smallest error (and its standard deviation) for the largest window size k = 10 (mm). The smaller the window, the less useful information network obtained, and the worse the quality of the estimation was. This observation is consistent with our initial assumption that the size of the input RF data can impact CNN-r's performance.

The average estimate values ($\pm$ standard deviation) for individual attenuation levels are presented in Fig. \ref{fig:res}. The larger the input size was, the closer the average estimate was to the true AC. A similar relation can be observed for the standard deviation. Moreover, the smaller the data size, the closer to AC of 0.5 the average estimate was. For example, for k = 10 (mm), the average estimate for AC = 0.1 and AC = 1.5 data was approx. 0.13 and 1.43; for k = 1 (mm) it was 0.43 and 1.1 respectively. Finally, it is important to note that the points in the Figure \ref{fig:res} arranged in a straight line -- that is, on average, the true order of AC values was retained by CNN-r.


\begin{table}
\centering
\begin{tabular}{|c|c|}
\hline
Window size & Average absolute error ($\pm$ std. dev.) \\ \hline \hline
10 mm       & 0.08 ($\pm$ 0.07)               \\ \hline
5 mm        & 0.12 ($\pm$ 0.11)             \\ \hline
2 mm        & 0.20 ($\pm$ 0.19)               \\ \hline
1 mm        & 0.25 ($\pm$ 0.22)                \\ \hline
\end{tabular}
\caption{Average error $|\hat{y_i} - y_i|$ ($\pm$ standard deviation) for several selected window sizes.}
\label{tbl:results}
\end{table}

\section{Conclusions}

In this work, we positively verified the feasibility of using convolutional neural networks to estimate ultrasound attenuation coefficient based on RF signal data. We presented for a simple two layer neural model that, after an appropriate number of training iterations, its weights can reflect expected loss of signal's spectra. In our experiments we noticed, that CNN's performance depended directly on the size of the input ultrasound data.

Our work can be extended in several ways. It would be interesting to verify what is the real-world case scenario performance of the neural network models trained using simulated RF data. The idea of preparing model on a large synthetic dataset and employing it to estimate AC for real data is very promising. This method may also help asses and improve ultrasound simulation software.    



\bibliographystyle{IEEEbib}
\bibliography{ref}

\end{document}